# Memristive integrative sensors for neuronal activity

Isha Gupta, Alexantrou Serb, Ali Khiat, Ralf Zeitler, Stefano Vassanelli, Themistoklis Prodromakis

**The advent of advanced neuronal interfaces offers great promise for linking brain functions to electronics. A major bottleneck in achieving this is real-time processing of big data that imposes excessive requirements on bandwidth, energy and computation capacity; limiting the overall number of bio-electronic links. Here, we present a novel monitoring system concept that exploits the intrinsic properties of memristors for processing neural information in real time. We demonstrate that the inherent voltage thresholds of solid-state TiOx memristors can be useful for discriminating significant neural activity, i.e. spiking events, from noise. When compared with a multi-dimensional, principal component feature space threshold detector, our system is capable of recording the majority of significant events, without resorting to computationally heavy off-line processing. We also show a memristive integrating sensing array that discriminates neuronal activity recorded in-vitro. We prove that information on spiking event amplitude is simultaneously transduced and stored as non-volatile resistive state transitions, allowing for more efficient data compression, demonstrating the memristors' potential for building scalable, yet energy-efficient on-node processors for big data.**

For decades, advances in microelectrode fabrication technology[1,2,3], power-efficient data acquisition and processing software[4], hardware[5,6,7,8] and systems have led to "faster, cheaper, better" neural recording systems capable of monitoring increasingly many electrode sites simultaneously[9,10]. Currently, state-of-art neural recording systems are capable of simultaneous recording from up to 512 sites at 40kS/s in-vivo[11] and from up to 32768 channels at 2.4kfps in cell-culture pixel arrays[12]. However, further advances are hindered by the increasing tightening of some key bottlenecks, notably increasing power dissipation (largely stemming from the cost of transmitting finely sampled, multi-channel data to a processing unit), dropping signal-to-noise ratio and routing considerations[13]. Addressing these challenges necessitates the intelligent compression of big data[14] via on-node processing, currently pursued by shifting the spike-sorting task on-chip[4,8,10]. However, the resulting scalability issues, the drive for even further power budget reductions and the fact that many neuroprosthetics can be successfully operated with simple, rate- or spike-count-coded input signals[5,15,16,17] (knowledge of precise spike timing no longer required) have kindled interest in processing extracellular potential data in a bio-inspired fashion. This justifies current interest in leveraging emerging technologies for resurrecting Carver Mead's original vision in neuromorphic systems[18], where efficient data processing is implemented through e.g. artificial retinas[19]. Memristive devices appear to be well-suited in providing a disruptive technological boost to this vision by performing the role of artificial synapses. Much akin to biological synapses, they possess the intrinsic ability to simultaneously carry out computational tasks and store information at aggressively downscaled volumes and power dissipation[20,21] . Here, we exploit intrinsic characteristics of TiOx memristors, such as input thresholding and analogue modulation of resistive state (RS), in order to encode neuronal spiking activity while suppressing noise. We demonstrate for the first time how a large part of the computational burden associated with spike processing can be relegated to highly scalable nanodevices that can be manufactured in the back-end-of-line part of Complementary Metal-Oxide Silicon (CMOS) electronics fabrication processes, thus costing potentially no chip layout area.

**Memristors as Integrating Sensors**

A corollary of Chua's original definition of the memristor[22] is that memristors are capable of changing their RS as a function of the integral of their input voltage (or current); a phenomenon known as 'resistive switching'. As a result of this 'single-device integrator' property, emerging solid-state devices exhibiting memristive behaviour[23,24,25] have been at the centre of attention, with potential applications in nanoelectronic memories and neuro-inspired computer hardware[26]. In this work we exploit metal-oxide-based resistive switches as integrating sensors. Solid-state $TiO_2$ memristors with vertical stack structure $Ti/Pt/TiO_{2-x}/Pt$ (5/10/25/10 nm) as shown in Fig. 1a were fabricated on a $Si/SiO_2$ substrate. All devices initially undergo an electroforming step[27] before use (Fig. 1b inset). A voltage sweep is applied on a pristine sample until a sudden, non-volatile change in RS is observed. This typically occurs at approximately +6.5V. Subsequently, bipolar switching can be observed with abrupt transitions between high RS (HRS) and low RS (LRS) occurring at around ±3V (Fig 1b). Subjecting the device under test (DUT) to a pulse-based stimulation protocol gives rise to gradual SET/RESET transitions. Figs. 1c and 1d show the response of a typical DUT to pulse trains consisting of 200 identical square-wave programming pulses of negative and positive polarities[28] respectively. Each programming pulse (fixed duration 100μs, ad hoc chosen amplitude) was followed by a non-invasive 'read' pulse (fixed amplitude 0.5V, automatically determined duration[29]) designed to provide an assessment of RS at low voltage and fixed polarity. The DUT supports bidirectionally gradual, saturating switching, that can be fitted by second order exponential functions of input voltage integral (magnetic flux); thus defining the input-output relation of an integrating sensor for the employed stimulation protocol. The required programming pulse magnitude is different for each polarity, indicating an inherent asymmetry in switching behaviour. It is this capability for gradual switching that allows individual memristors to encode multiple, significant events in its input history as small changes in its RS.

Importantly, many families of practical RRAM devices, including our TiOx-based samples, possess an intrinsic voltage threshold below which the effects of pulsing on RS are negligible. Correction factors may be easily introduced in the second order exponential fitting functions in order to account for the effects of such thresholds and yield the same fitted curve (see supporting material). Such devices act as thresholded integrators (Fig. 2a) and are better described by the generalised definition of memristors as "zero phase shift dynamic systems" [30]. This behaviour is apparent in Figs. 2b,c where a DUT was subjected to progressively more invasive trains of programming pulses in alternating polarities. The RS was read after each programming pulse. Significant changes in RS are observed in Fig. 2c, only after stimulus pulse voltages exceed the positive and negative polarity intrinsic thresholds of the DUT, here identified as $V_{th+}$ = 1.45V and $V_{th-}$ = -1.65V respectively. All of our TiOx devices exhibit similar threshold voltages, even though the precise values vary between them (supplementary Fig. S2). This attribute is useful for suppressing noise in signals with low signal-to-noise ratios (SNR), such as data recorded from neuronal activity. Only significant, supra-threshold events in the input waveform will be registered as measurable changes in RS. Figure 2d shows a test waveform consisting of four, concatenated copies of a neural recording. The recordings were obtained from a Multi Transistor Array (MTA) CMOS chip[2,31,32] and PCB-mounted trans-impedance amplifiers (TIA) that boost the signal up from the 0.1-1 mV range (see methods section). Each copy was subjected to scaling and offsetting, and two of the copies were polarity-inverted. The waveform was then fed into a different device in batches of 1000 data-points (Fig.S3). RS is assessed at the beginning of each batch, then every 300 samples and at the end of the batch (standard scheme: assess initial RS and after application of the $300^{th}$, $600^{th}$, $900^{th}$ and $1000^{th}$ data-points). Offset and scaling parameters were chosen so that only a small number of the most significant events would exceed $V_{th+}$ and $V_{th-}$. The evolution of DUT RS throughout the test is shown in Fig. 2e. Significant changes in RS correspond to clear, supra-threshold spiking events, spike detection successfully occurs in both polarities and the results are repeatable over two signal periods. This is better illustrated in Fig. 2f, where the normalised RS change between consecutive 'reads' is plotted as a function of the maximum voltage magnitude of interceding events (Fig. 2d). One can observe three distinct areas: two of them corresponding to significant RS modulation (above $V_{th+}$ and below $V_{th-}$) and the last one ($\in[V_{th+}, V_{th-}]$) containing RS changes indistinguishable from background noise. An estimate of background noise can be extracted by logging the changes in RS between the final measurement of each batch and the initial measurement of the next one. Additional measurements shown in Fig. S2 validate this concept over multiple neural recordings that cover a wide variety of activity patterns, i.e. number and density of events.

**A memristor-based neural activity sensor.**

The ability of memristors to detect significant events in neural recording waveforms presents an efficient way of encoding and processing neuronal activity. The proposed system concept is illustrated in Fig. 3a. Neuron activity is sensed through a front-end circuit as analogue current or voltage and subsequently conditioned through a linear gain stage followed by a constant offset stage before biasing the DUT. This offers the flexibility of tuning the SNR of our memristive integrating sensor, ensuring tailored monitoring of events whilst suppressing noise. Periodically, the memristor is disconnected from the neural signal feed and connected to a read-out circuit, which assesses its RS, digitises it and therefore only sends a limited amount of data (compared to the full voltage time series) off-chip. Spike information is then extracted by thresholding the resulting changes in RS (ΔRS) between consecutive measurements. While the intrinsic threshold of the devices performs a coarse filtering action, the value of this second, ΔRS threshold may be arbitrarily chosen for fine-tuning SNR. Further data post-processing may be employed in order to set the ΔRS thresholds adaptively, however in this work we concentrate only on system operation up to basic spike detection results. Our efforts to reduce data bandwidth echo current research in on-chip spike-sorting[8], but our approach is advantageous as it exploits highly scalable, low power nanodevices that could extend the scaling capacity of integrated neural recording platforms substantially. In contrast, conventional neural activity monitoring platforms like the MTA-based system described in [31] and illustrated in Fig. 3b, utilise front-end circuitry for detecting and transmitting all data to a computer that performs offline processing. In the case of the MTA-based system above, this consists of principal component analysis on a 27-parameter kernel covering a 3x3x3 spatial/spatial/temporal cube followed by thresholding. Supra-threshold peaks are extracted as spiking events. Similar to our system, additional data post-processing may set these thresholds adaptively.

We validated our approach (Fig. 3a) experimentally with sample traces of spontaneous retinal activity, recorded with the front-end of the MTA-based CMOS system in[31] (see methods section) at a 12.2kfps sampling rate. The front-end consists of the MTA itself, which outputs a raw current time-series (approximately 64kS), and board-mounted TIAs which convert and amplify the signal to voltage in the hundreds of mV range. The conditioned signal is then fed to an in-house developed memristor characterisation instrument[33,29] (methods section and supplementary information). The instrument applies offset ($V_{ofs}$ = 0) and amplification (G = 2.8) before playing the conditioned recording, shown in Fig.3c, to the DUT in 1kS batches, with RS assessed as per the standard scheme and recording play-back suspended during RS read operations. Figure 3d illustrates the RS evolution of our memristive integrating sensor in response to the input signal of Fig. 3c. One can observe clear changes in RS corresponding to spiking events whose magnitude exceeds $V_{th-}$, in a similar manner to the first period of events in Figs. 2d,e. In this example, the incoming spikes dominate for the negative polarity hence there is an overall increase in DUT RS, from 2.5 kΩ to 5.5kΩ. However, the presence of a few events in opposite direction that exceed $V_{th+}$ often cause a decrease in RS. This can be observed at approximately 1.4s (more clearly visible in Figs. 3g,h) where the RS drops from 4.5kΩ to 3kΩ. Optimising the value of $V_{ofs}$ is expected to be a good way to compensate for this effect. Employing the processing schematic of Fig. 3b we were able to discriminate 81 significant events (shown Fig. 3e) within the recording of Fig. 3c. Since our memristive integrating sensor is capturing and storing significant events as non-volatile changes in RS we can afford to utilise relatively low sampling rates at the expense of timing resolution. As a result, the output of our system is quantified as distinct time bins containing one or more significant events. Such a case arises at approximately the 0.96s mark and can be seen in detail in Figs. 3g-j. Results through the proposed (Fig. 3a) and conventional approaches (Fig. 3b) concur for the majority of events, as indicated via '*' marks in Fig. 3j. However, it is interesting to note that the integrating sensor registers apparent events (e.g. 0.83s, 1.05s, 1.1s) that are missed by the conventional method while missing other possible events (e.g. 0.59s, 1.22s, 1.25s), presumably due to a conservative selection of gain. As a result, our system identifies 74 bins overall, shown in Fig. 3f, as containing significant events.

The concept introduced in Fig. 3a is amenable for scaling to the multi-channel array level, as illustrated in Fig. 4a. We envisage an overall system architecture very similar to standard, Active Pixel Sensor (APS) CMOS imagers[34]. Data from each pixel, arrives as an analogue current from the MTA and is multiplexed onto an on-chip TIA block, which is followed by an on-chip offset stage. Thus, both gain and offset stages are time-shared by every pixel in the array. The conditioned recording data points are then de-multiplexed to a memristor bank, integrated into the back-end of the chip, close to the MTA recording sites. Memristor read-outs, occurring at a data-rate reduced by a factor of 200 compared to the raw voltage time series if the standard biasing scheme is followed, would be processed sequentially from each memristive device through a time-shared Analogue-to-Digital Converter (ADC) and then sent off-chip. The working principle is validated by using our system from Fig. 3a to process 224 distinct neural recording traces originating from a 16x14 pixel sub-array in an MTA system[2,7] atop which retinal cells have been cultivated. The sub-array covers three retinal cells (supplementary Fig. S5). An initial calibration was performed in order to set suitable values for G and $V_{ofs}$. This entailed selecting a spatially sparse subset of 23 pixels (see supplementary Fig. S4, cells marked in orange) and examining their neural recording waveforms in order to gauge average maximum/minimum voltage amplitudes as well as the typical levels of background activity. In this experiment, we set G = 2.8 and $V_{ofs}$ = 0 to ensure a useful SNR for our memristive integrating sensing array. These parameters were kept fixed for all recordings, i.e. they were not changed in order to accommodate either individual memristive device behavioural variations, or the specific characteristics of individual recordings (e.g. presence of unusually high amplitude spikes, relative spiking event level vs. baseline activity etc.). Every employed memristive device was initialised to a common LRS in the range of 2-4kΩ that for the given parameters yielded a useful operating range up to the 15kΩ HRS. In turn, we obtained a set of 224 RS time evolution traces similar to the one shown in Fig. 3d. Snapshots of the array state as RS(t)/RS($t_0$) are shown in Figs. 4b,c,d, at $t_1$ = 1.63s, $t_2$ = 3.27s and $t_3$ = 5.16s respectively. Since neuronal activity is encoded as non-volatile RS changes we are able to observe an accumulation of activity clustered around three major centres at pixel (row, column) locations (3,4), (7,10) and (11,7). Particularly the final array state qualitatively resembles the whole extracted activity, shown in S5, for the same neural recording data-set as obtained by a state-of-art array-level system operating on the principle shown in Fig. 3b. We note that whilst the system in S5 outputs a spike count, insensitive to the amplitude of the detected spikes, the memristive integrating sensing array outputs a ratiometric change in RS that is strongly correlated to the strength of the individual spiking events. This allows us to preserve information on event amplitude and even polarity, which in principle improves data compression rate. We also note a few pixels exhibiting strong RS changes despite not appearing to belong to any well-defined cluster of activity (see supplementary Fig. S6). This discrepancy follows the argument presented previously for Figs. 3i-j, hinting that single, exceedingly strong events may lead to RS changes comparable to those arising as a result of accumulated activity.

**Summary**

We have demonstrated a novel neural recording system concept that addresses challenges associated with the processing of large datasets resulting from high spatio-termporal resolution neural activity sampling. Our approach directly exploits the intrinsic 'synapse-like' attributes of emerging memristor prototypes, in particular the capacity of TiOx-based memristors to integrate input signals above certain threshold limits; a significant stepping stone towards the development of bio-inspired neural signal processing system. Results show that single devices are capable of identifying significant spiking events while suppressing noise, thus paving the way towards highly area- and energy-efficient on-node neural recording processing. Contrary to time-domain sampling, both conventional and emerging techniques (e.g. address-event representation), the proposed memristive integrating sensors encode the presence of events in non-volatile RS changes, allowing the flexibility to trade off sampling rates for timing resolution. This is particularly useful in applications where data is rate- or spike-count-coded and therefore only a measure of overall activity within given time bins is required, as is the case in numerous neuroprosthetic applications. Moreover, as memristive device RS changes are linked to input signal strength, much of the richness of information within each time bin is preserved in the magnitude of RS modulation, albeit distorted by cycle-to-cycle variations in switching behaviour. With improving fabrication techniques and switching stability, this property could become particularly useful for in-vivo recordings, where the amplitude of events provides an indication of spatial proximity between the spiking cell and the recording electrode. Finally, this concept can be further applied for ‘smart’ data compression in distinct sensing platforms, particularly relevant to pervasive sensing systems.

**Figure 1** Device architecture and electrical characterisation of solid-state $TiO_2$ ReRAM devices. (a) Schematic illustration of a TiOx RRAM memristor. (b) DC characterization of the electroformed devices. Voltages are applied to the top electrode (TE) whilst the bottom electrode (BE) is grounded for all the measurements. The inset shows the electroforming step where the device forms to ‘ON’ state under positive bias. After electroforming the device exhibits bipolar switching with low resistive state (LRS) to high resistive state (HRS)

transitions. RESET occur under negative bias whilst the complementary SET transitions occur under positive bias. (c) and (d) show gradual resistive switching under pulse train input stimulation (200 pulses per train). Device responses are fitted with second order exponential functions. Typical biasing scheme parameters (insets): positive write pulse voltage $V_{w+}$ = +0.8V, negative write pulse voltage $V_{w-}$ = -1.2V, read pulse voltage $V_a$ = 0.5V, write pulse width $t_w$ = 100µs and read pulse width $t_a$ automatically determined by the measurement system.

**Figure 2** A practical memristive device typically performs the function of a thresholded input integrator. (a) Conceptual behavioural model of a thresholded input integrator that resembles a memristive device. This contains a dual threshold ($V_{th+}$/$V_{th-}$) detection block which controls whether the input is passed to the integrator block (∫) or not. This behaviour becomes apparent when a device is biased as with voltage pulse trains of increasing amplitudes as shown in (b). (c) depicts the corresponding RS changes accumulating visibly only in response to input pulses with above-threshold amplitudes. In bipolar devices there exist two voltage thresholds, one for each voltage polarity. For this device we obtained Vth+ = +1.45V and Vth- = -1.65V. (d) Shows an arbitrary input waveform consisting of four copies of a retinal cell neural recording in alternating polarities. The recording contains multiple spiking events. This waveform was employed to validate the concept of memristive integrating sensors, the result of which is shown in (e). The collated recording copies in (d) have been subjected to appropriate scaling and offsetting in order to accommodate the device's asymmetric threshold voltages, identified as $V_{th+}$ = 1.2V and $V_{th-}$ = -1.5V; resulting in balanced RS SET and RESET. x-axis for both (d) and (e) is given in S.I. units – each data sample lasts 82µs. (f) Fractional RS modulation ($\Delta R/R_o$) extrapolated from (e) showcasing significant RS modulation occurring only above $V_{th+}$ and below $V_{th-}$ while intermediate bias values (noise) leads to no significant change.

**Figure3** Signal conditioning in the memristive and the template-matching system. (a) Schematic illustration of proposed memristive integrating sensor signal processing. A front end system extracts recordings from the cell culture which are then suitably gain-boosted (G) and offset ($V_{ofs}$) to render them compatible with the memristors' voltage operating regimes. Changes in RS due to significant spiking events can be extracted offline. (b) Template matching-based spike detection concept. (c) illustrates a pre-conditioned neural recording trace that causes the RS time evolution shown in (d). (e) 81 spikes were detected by the template matching system, with grey lines indicating spike positions. (f) Green shading indicates time intervals within which one or more spikes were detected through the memristive integrating sensor; total of 74. (g) and (h) are close-ups of the neural recording and RS evolution shaded grey in (c) and (d) respectively. Time intervals where the memristive integrating sensor detects spikes are shaded green whilst the locations of spikes detected by the template matching system are indicated by vertical dashed lines. (i) and (j) Comparison of the detection of spikes by two systems. The asterisk mark indicates the instances of agreement between the two systems.

**Figure 4** A memristive integrating sensing array. (a) Conceptual diagram indicating conditioning of multi-channel data through a common gain and offset cascade. (b), (c) and (d) Time evolution of normalised RS throughout 16x14 test array at $t_1$ = 1.63s, $t_2$ = 3.27s and $t_3$ = 5.16s respectively. Three clusters of elevated activity can be discriminated. CMOS MTA: Multi-transistor Array Block, manufactured in standard, commercially available CMOS

technology. TIA: Trans-impedance amplifier converting current to voltage with appropriate amplification.

**Methods Summary**

**Fabrication -** All the devices exploited in this work were fabricated according the following flowchart; 200 nm of insulating $SiO_2$ was thermally grown on 6-inch Silicon wafer. Then three main patterning steps were processed, each contains optical lithography, film deposition and lift-off process. In the first step, 5 nm Titanium (Ti) and 10 nm Platinum (Pt) films were deposited via electron-gun evaporation technology to serve as bottom electrodes, Ti was used for adhesion purposes. In the second, magnetron reactive sputtering system was used to deposit the $TiO_{2-x}$ (x=0.06) active core from Ti metal target. Two plasma sources were used to ensure homogeneous near stoichiometric film. 25 nm thick $TiO_{2-x}$ was deposited. In the final step, 10 nm Pt top electrodes were deposited using electron-gun evaporation system. At the end of processing, the wafer was diced into 9x9 $mm^2$ chips, which were then wire-bonded in standard packages for measurements.

**Hardware Infrastructure** - The biasing protocol was implemented using custom made hardware developed in-house. It consists of a microcontroller-based PCB-mounted system[33] capable of addressing devices embedded in crossbar arrays of up to 1kb in size (32x32). The system has the capability of either testing packaged arrays or communicating to a multi-channel probe card for direct testing on the wafer. The hardware is supported by custom-made software that permits exhaustive, device-by-device testing of entire crossbar array or an array of individual devices in one, fully automated round of measurements. The biasing schemes applied for read and write operations are the '$V_r$' (Fig.3[31]) and '$V_{r/2}$' (Fig.10b[32]) schemes, described in detail in their respective references. This helps in mitigating the sneak path effects.

**Mathematical Model** - For the DUT, curve-fitting was carried out using standard curve-fitting tool in MATLAB. The data from the RS of the devices for negative (Fig.1c) and positive (Fig.1d) pulses were separately fitted to second-order exponential function i.e $f(\int Vdt) = Ae^{\beta\int Vdt} + Be^{\Upsilon\int Vdt}$, where V is the fixed pulse voltage. The data for the mathematical model is tabulated in supplementary information (Table S1).

**Cell culture –** Neural activity from the portions of mid-peripheral rabbit retina was recorded using MTA fabricated in CMOS technology[37,38]. The preparation of the retina culture preserves its functional integrity. The surface of CMOS multi-transistor array comprising of 128x128 sensor sites is insulated by a thin, inert $TiO_2/ZrO_2$ layer. The oxide layer is connected to the gate of the field-effect-transistor via metallic pathway. The electronic current between the source and drain in the silicon-based field effect transistor is modulated by the application of local voltage changes within the interfaced neural tissue above the recording sites. The neural tissue is separated from the MTA oxide by an electrolyte-filled cleft.

**Supplementary Information** is available in the online version of the paper.

**Acknowledgements** We acknowledge the financial support of FP7 RAMP and EPSRC EP/K017829/1.

**Author contributions** T.P. and S.V. conceived the experiments. A.K. fabricated the samples. I.G.* and A.S.* performed the electrical characterisation of the samples and developed the control instrumentation and software. S.V. carried out the cell culturing and R.Z. developed the front-end recording platform. All authors contributed in the analysis of the results and in writing the manuscript.

*These authors contributed equally to the work.

**Author information** Reprints and permissions information is available at. The authors declare no competing financial interests. Correspondence and requests for materials should be addressed to T.P. (t.prodromakis@soton.ac.uk)

## Figure 1

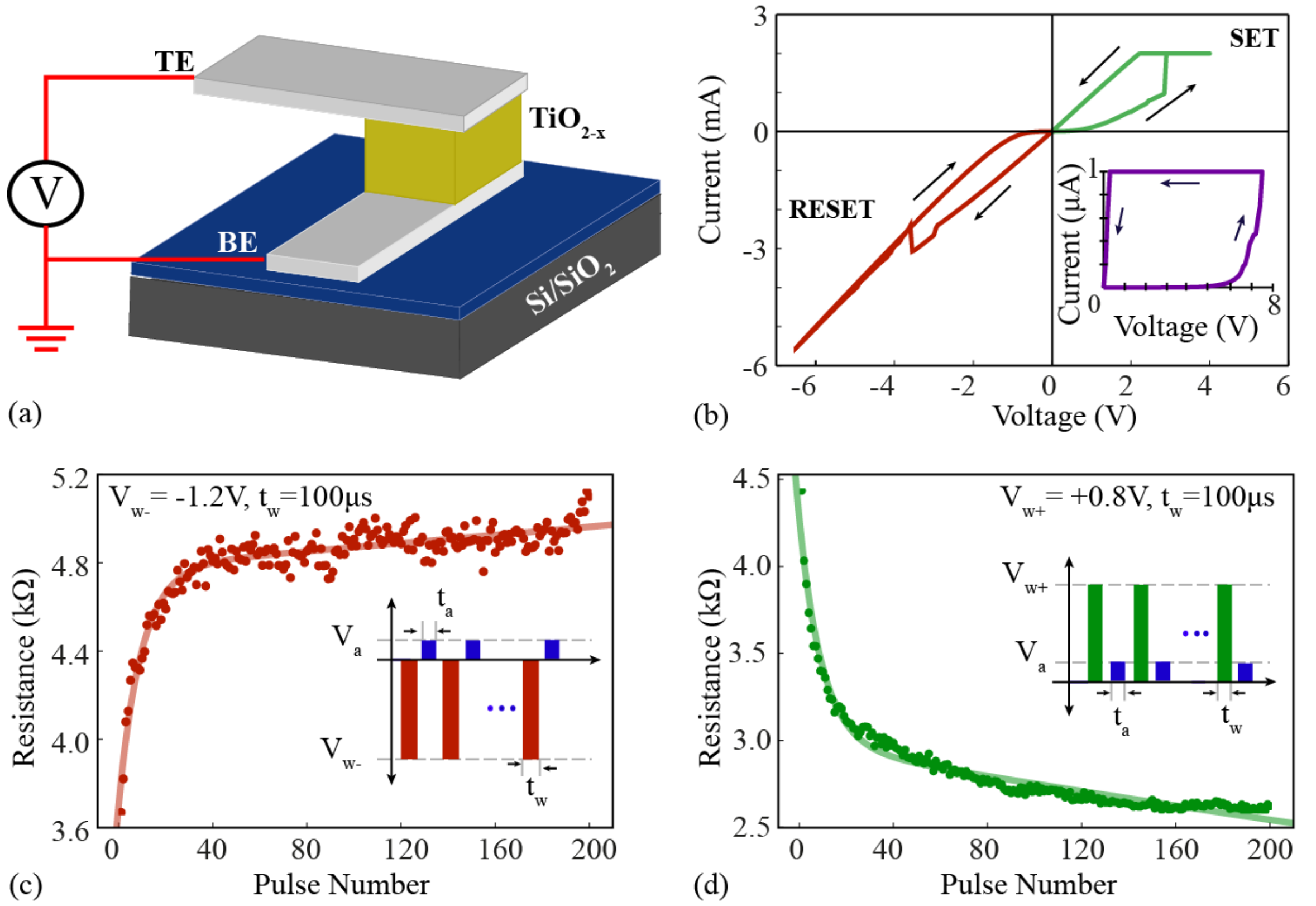

## Figure 2

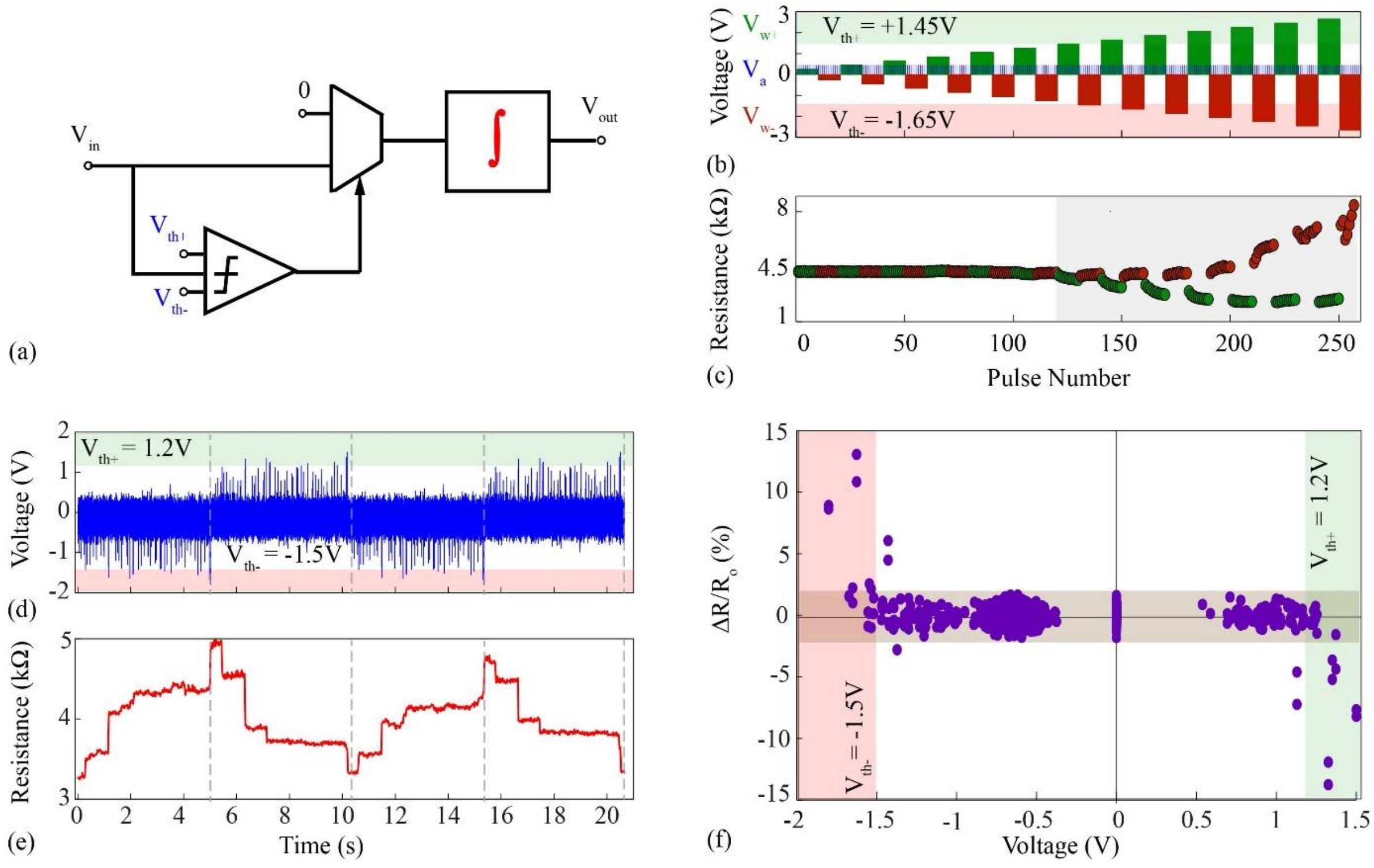

## Figure 3

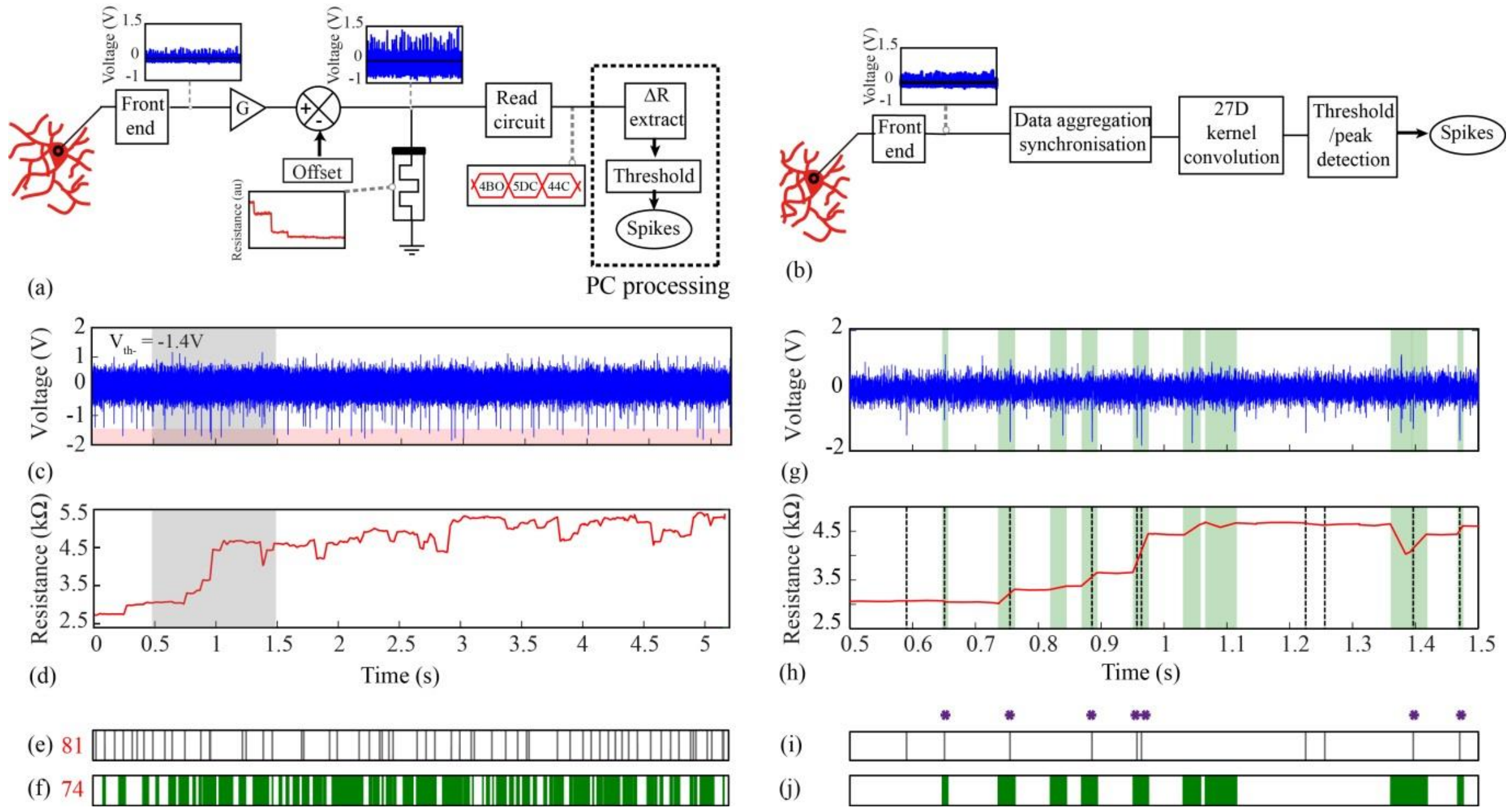

## Figure 4

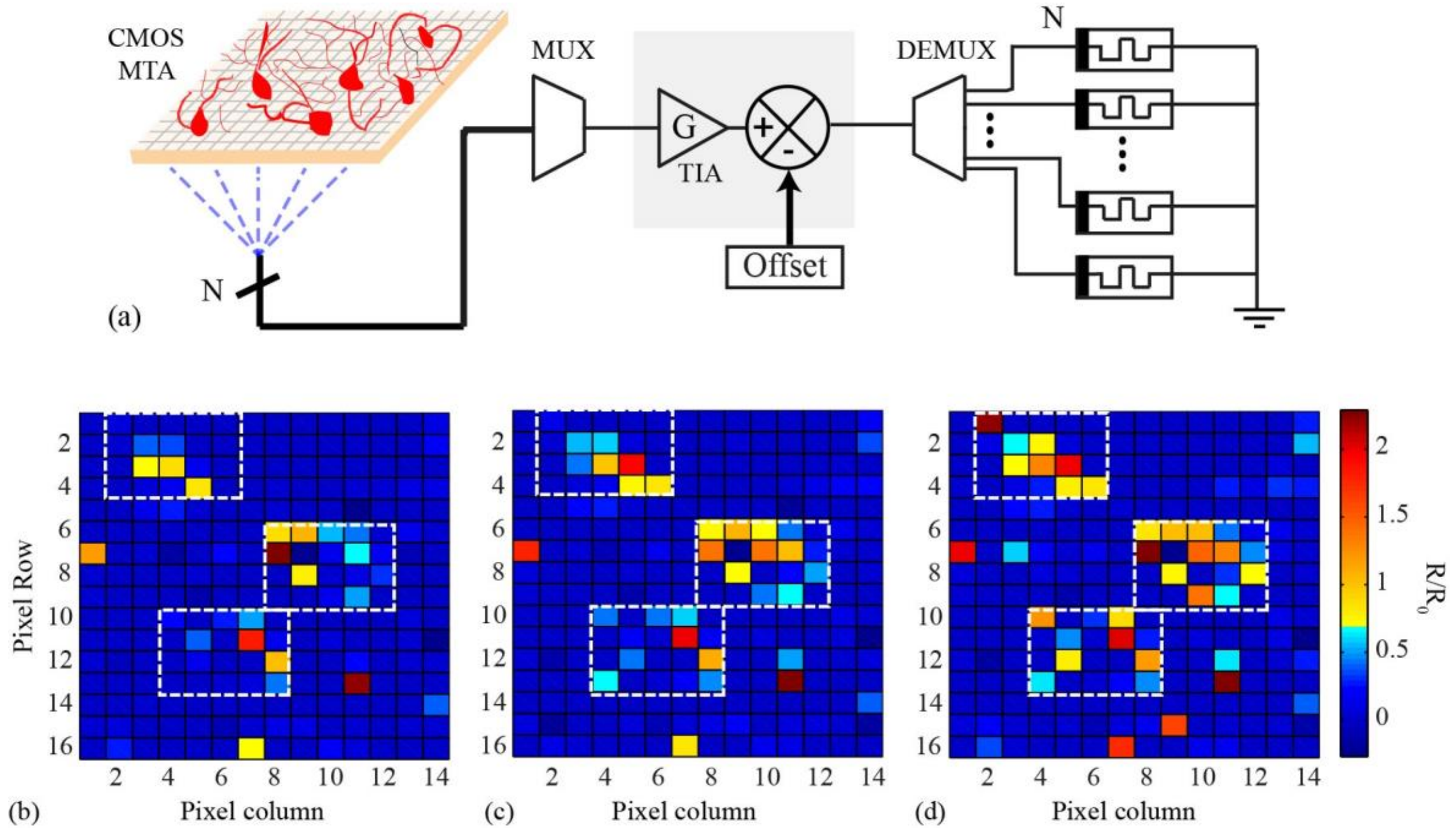

## SUPPLEMENTARY INFORMATION

# Memristive integrative sensors for neuronal activity

Isha Gupta, Alexantrou Serb, Ali Khiat, Ralf Zeitler, Stefano Vassanelli, Themistoklis Prodromakis.

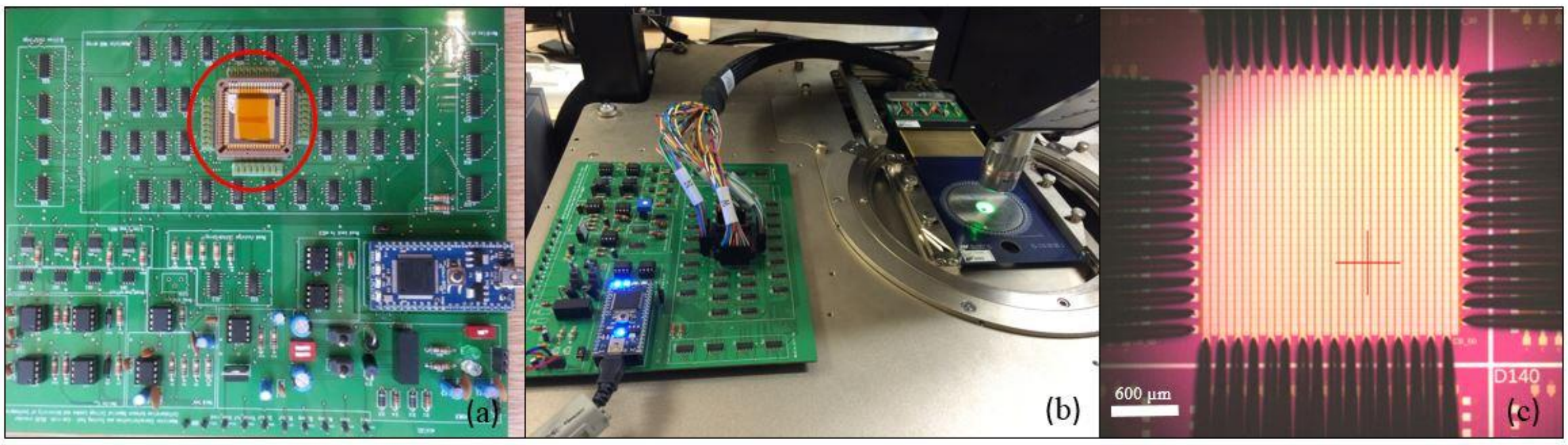

**Figure S1** Hardware Infrastructure used in order to implement proposed biasing protocol[1,2,3]. (a) PCB-mounted system overview with visible crossbar package (circled in red). (b) System connected to multi-channel probe card. (c) Microscope image of 32x32 crossbar showing probe card needles touchdown.

| Fitting coefficients (95% confidence upper/lower bounds) | | |
|---|---|---|
| Coefficient | Fig.1c | Fig.1d |
| A | 4780 (4756, 4803) | 1301 (1233, 1368) |
| β | $1.878e^{-4}$ ($1.492e^{-4}$, $2.264e^{-4}$) | $-1.027e^{-1}$ ($-1.118e^{-1}$, $-9.366e^{-2}$) |
| B | -1075 (-1156, -994.7) | 2976 (2954,2998) |
| ϒ | $-1.04e^{-1}$ ($-1.17e^{-1}$, $-9.098e^{-2}$) | $-7.829e^{-4}$ ($-8.418e^{-4}$, $- 7.24e^{-4}$) |
| Goodness of fit: | | |
| R-square | 0.9137 | 0.9701 |
| Adjusted R-square | 0.9124 | 0.9697 |
| RMSE | 56.77 | 47.8 |

**Table S1** The DUT responses shown in Fig.1c and 1d were fitted with second order exponential functions of input voltage integral taking the form $f(\int Vdt) = Ae^{\beta\int Vdt} + Be^{\Upsilon\int Vdt}$, where V is the fixed pulse voltage, using standard MATLAB curve-fitting tool. However under constant voltage pulse stimulation results yielded by the non-thresholded model can be exactly reproduced by a thresholded model of the form $f(x) = Ae^{\beta'(\int max(|V-V_{th}|,0)\ dt)} + Be^{\Upsilon'(\int max|V-V_{th}|,0)\ dt}$ where $\beta' = \beta|V/(V-V_{th})|$, $\Upsilon' = \Upsilon|V/(V-V_{th})|$. This holds for $|V|>|V_{th}|$. Statistical parameters extracted from curve-fitting are summarised in the table above. The fitted curve is in good agreement with the obtained experimental data.

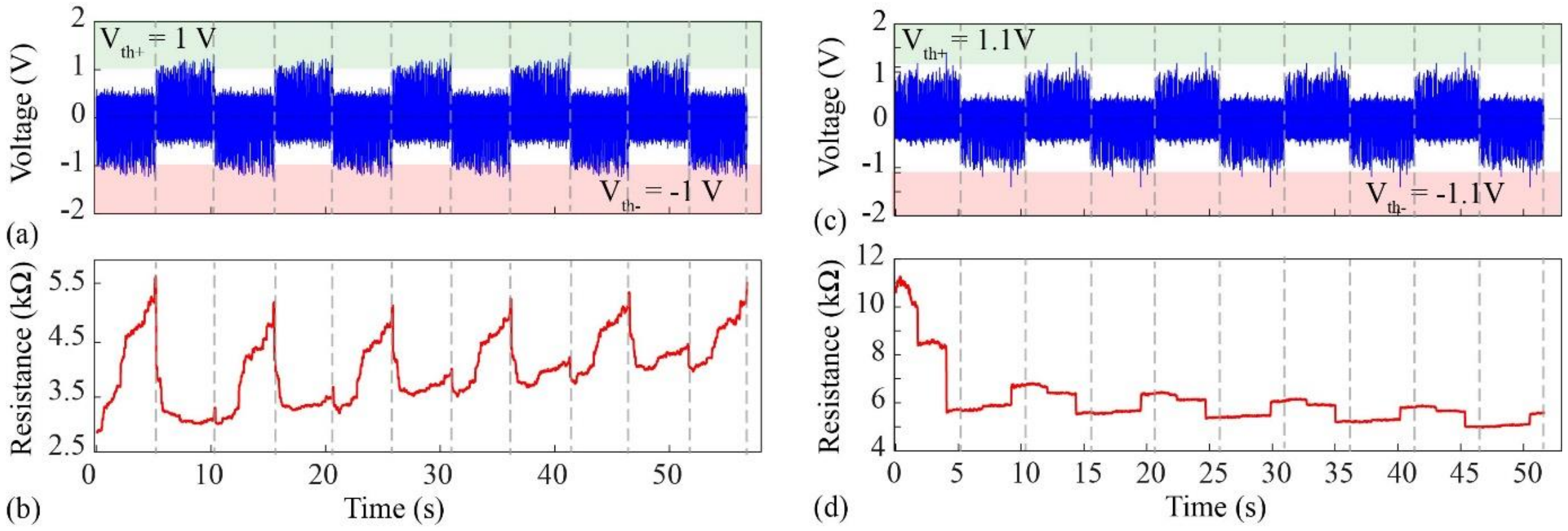

**Figure S2** Activity patterns of two DUTs when subjected to repeated instances of a neural recording with alternating polarities. The devices clearly illustrate qualitatively repeatable behaviour over a number of cycles which indicates a measure of robustness to switching variation. (a) Multiple cycles of a neural recording are concatenated in alternating polarities and fed to a DUT. Grey vertical dashed lines demarcate the different copies of the recording. (b) DUT response. The threshold voltages ($V_{th+}$ and $V_{th-}$) of the device are ±1V. An overall drift is observed in this device, illustrating the effects of cycle-to-cycle variations in switching behaviour. (c),(d) Similarly, multiple cycles of another recording are fed to the second DUT, which yields a gradually stabilising response. Threshold voltages: ±1.1V.

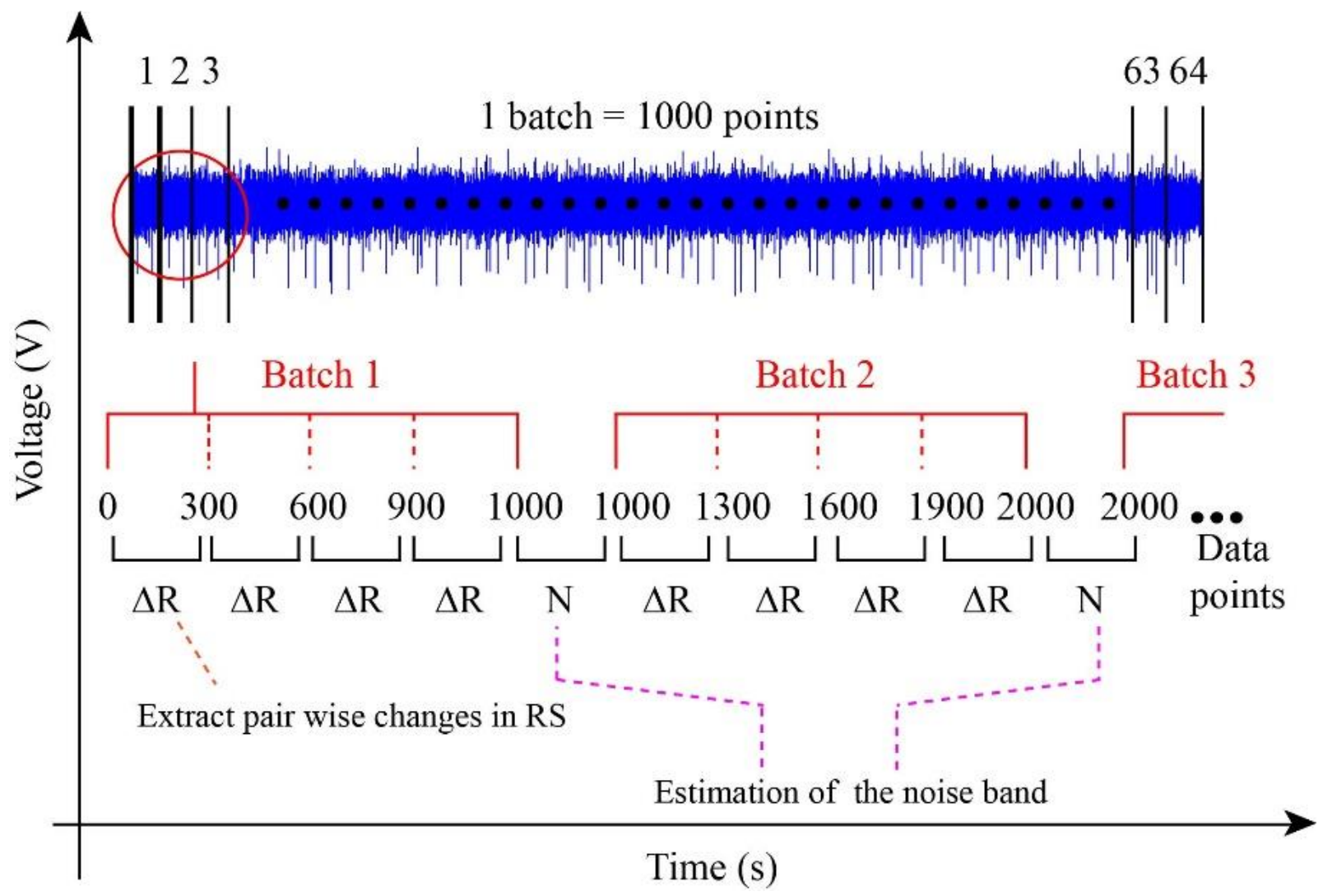

**Figure S3** Standard read-out scheme for evaluating the time evolution of the resistive state (RS) of test devices subjected to input stimulation. Input data is fed to each device in batches of 1000 points with RS assessed at the beginning of each batch, then every 300 samples and finally at the end of each batch. Changes in test device RS (ΔR) are extracted from consecutive RS assessments. RS changes occurring between the last measurement of each batch and the first measurement of the next batch, i.e. with no interceding pulse biasing, (N) are considered to result from measurement uncertainty and can be used to determine the noise band.

| S1_Cx_Ry | C30 | C32 | C34 | C36 | C38 | C40 | C42 | C44 | C46 | C48 | C50 | C52 | C54 | C56 |
|---|---|---|---|---|---|---|---|---|---|---|---|---|---|---|
| R82 | 1 | 17 | 33 | 49 | 65 | 81 | 97 | 113 | 129 | 145 | 161 | 177 | 193 | 209 |
| R83 | 2 | 18 | 34 | 50 | 66 | 82 | 98 | 114 | 130 | 146 | 162 | 178 | 194 | 210 |
| R84 | 3 | 19 | 35 | 51 | 67 | 83 | 99 | 115 | 131 | 147 | 163 | 179 | 195 | 211 |
| R85 | 4 | 20 | 36 | 52 | 68 | 84 | 100 | 116 | 132 | 148 | 164 | 180 | 196 | 212 |
| R86 | 5 | 21 | 37 | 53 | 69 | 85 | 101 | 117 | 133 | 149 | 165 | 181 | 197 | 213 |
| R87 | 6 | 22 | 38 | 54 | 70 | 86 | 102 | 118 | 134 | 150 | 166 | 182 | 198 | 214 |
| R88 | 7 | 23 | 39 | 55 | 71 | 87 | 103 | 119 | 135 | 151 | 167 | 183 | 199 | 215 |
| R89 | 8 | 24 | 40 | 56 | 72 | 88 | 104 | 120 | 136 | 152 | 168 | 184 | 200 | 216 |
| R90 | 9 | 25 | 41 | 57 | 73 | 89 | 105 | 121 | 137 | 153 | 169 | 185 | 201 | 217 |
| R91 | 10 | 26 | 42 | 58 | 74 | 90 | 106 | 122 | 138 | 154 | 170 | 186 | 202 | 218 |
| R92 | 11 | 27 | 43 | 59 | 75 | 91 | 107 | 123 | 139 | 155 | 171 | 187 | 203 | 219 |
| R93 | 12 | 28 | 44 | 60 | 76 | 92 | 108 | 124 | 140 | 156 | 172 | 188 | 204 | 220 |
| R94 | 13 | 29 | 45 | 61 | 77 | 93 | 109 | 125 | 141 | 157 | 173 | 189 | 205 | 221 |
| R95 | 14 | 30 | 46 | 62 | 78 | 94 | 110 | 126 | 142 | 158 | 174 | 190 | 206 | 222 |
| R96 | 15 | 31 | 47 | 63 | 79 | 95 | 111 | 127 | 143 | 159 | 175 | 191 | 207 | 223 |
| R97 | 16 | 32 | 48 | 64 | 80 | 96 | 112 | 128 | 144 | 160 | 176 | 192 | 208 | 224 |

Row ID

Column ID

**Figure S4** Overview of the MTA 16x14 sub-array recordings. Green cells: MTA column coordinates of recordings employed. Orange cells: MTA row coordinates of recordings employed. Blue cells: unique pixel identifier numbers. Dark blue cells indicate recordings that contain at least 1 data sample with magnitude above 0.5V; this includes a number of recordings that barely pass the 0.5V mark. The map therefore represents the presence of spikes as benchmarked by simple threshold method. The 23 recordings shown in orange colour were chosen as a representative sample of the entire array and were then used to estimate gain (G) and offset values ($V_{off}$) for the main experiments.

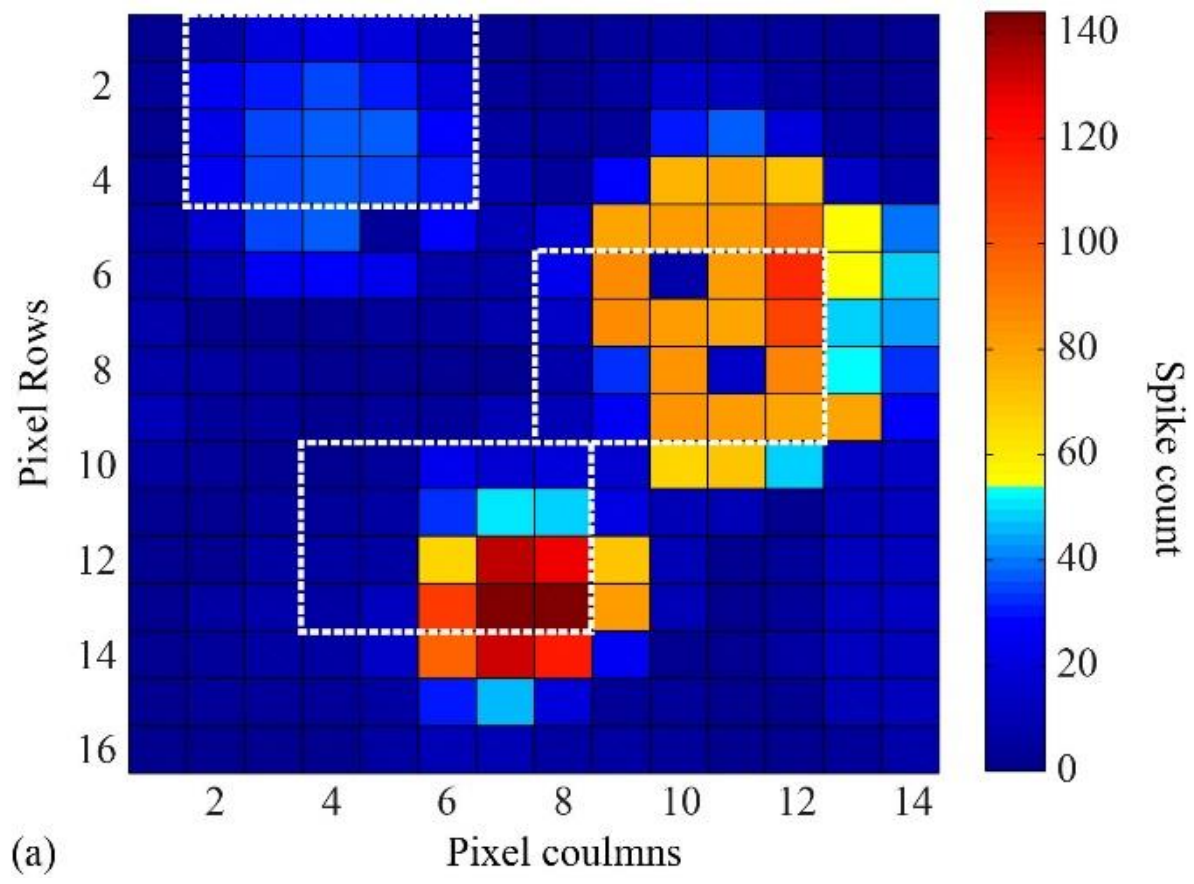

**Figure S5** Spike count pixel map yielded by the conventional system[4,5] in for the MTA sub-array recordings used to obtain Fig. 4 in the main text. The conventional system employs standard principle component analysis method (PCA) to extract spikes. Three major regions of activity can be clearly seen, which roughly correspond to the three main regions of activity detected by the memristive system (marked by white, dashed lines). Colour bar mapping has been chosen to correspond to the one employed in main text Fig. 4.

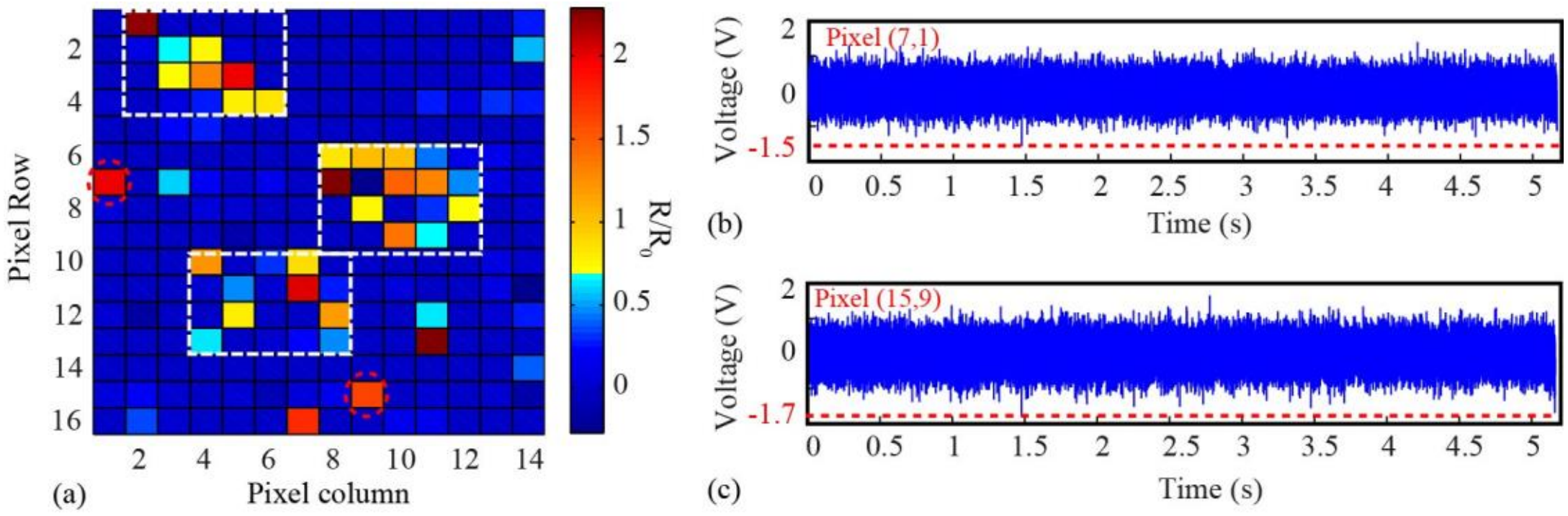

**Figure S6** The origin of 'stray' pixels. (a) Memristive device array final state pixel map reproduced from main text Fig. 4d. It shows the DUT final/initial RS state ratio. The three main regions of activity are enclosed in white, dashed line boxes, as are two of the four 'stray' pixels that show remarkable change in state without belonging to any of the three main clusters. (b) and (c) We investigated the neural recordings corresponding to these stray pixels and found that they all contain lone, remarkably high amplitude events; examples from two such recordings are circled in red. The stray pixels therefore react to real events in the stimulus waveform.